# DATA AGGREGATION AND PRIVACY FOR POLICE PATROLS

Sumalatha Ramachandran[1], Uttara Sridhar[2], Vidhya Srinivasan[3], J. Jaya Jothi[4]

[1-4]Department of Information Technology, MIT Campus, Anna University, Chennai - 600044, India

[1]sumalatha.ramachandran@gmail.com
[2]uttara.26790@gmail.com
[3]vidhya.s89@gmail.com
[4]jjothi42@yahoo.com

## ABSTRACT

*With a widespread growth in the potential applications of Wireless Sensor Networks, the need for reliable security mechanisms for them has increased manifold. This paper proposes a scheme, Privacy for Police Patrols (PPP), to provide secure data aggregation that relies on multilevel routing. Privacy factors have been identified and implemented. Aggregates are prepared and the summary of information is gathered and stored in a repository. The above defined approaches are integrated in police patrol applications and preliminary results are obtained.*

## KEYWORDS

*WSN, location privacy, secure routing, context summarization, aggregation, police patrols*

## 1. INTRODUCTION

A Wireless Sensor Network (WSN) consists of spatially distributed autonomous sensors to cooperatively monitor physical or environmental conditions, such as temperature, sound, vibration, pressure, motion or pollutants. A sensor network normally constitutes a wireless ad-hoc network that is each sensor supports a multi-hop routing algorithm where nodes function as forwarders, relaying data packets to a base station.

Unlike traditional wireless devices, wireless sensor nodes do not need to communicate directly with the nearest high-power control tower or base station, but only with their local peers. Instead, of relying on a pre-deployed infrastructure, each individual sensor or actuator becomes part of the overall infrastructure. Peer-to-peer networking protocols provide a mesh-like interconnect to shuttle data between the thousands of tiny embedded devices in a multi-hop fashion. The flexible mesh architectures envisioned dynamically adapt to support introduction of new nodes or expand to cover a larger geographic region. Additionally, the system can automatically adapt to compensate for node failures.

The resource-starved nature of sensor networks poses great challenges for security. However, in many applications the security aspects are as important as performance and low energy consumption. The security challenges include the extremely large number of interacting devices in a sensor network and the dynamic nature of WSN, that is, frequent changes in both its topology and its membership. Privacy is the ability of an individual or group to seclude them or information about themselves and thereby reveal who they are selectively. As location tracking capabilities of mobile devices are increasing, problems related to user privacy arise, since user's position and preferences constitute personal information and improper use of them violates user's privacy.

Wireless sensor networks are formed by small devices communicating over wireless links without using a fixed networked infrastructure. Because of limited transmission range, communication between any two devices requires collaborating intermediate forwarding network nodes, i.e. devices act as routers and end systems at the same time. Communication between any two nodes may be trivially based on simply flooding the entire network. However, more elaborate routing algorithms are essential for the applicability of such wireless networks, since energy has to be conserved in low powered devices and wireless communication always leads to increased energy consumption.

Data aggregation is an important primitive in wireless sensor networks (WSN). Data aggregation is a process that collects data from different sources and expresses the data in a summarized format. By eliminating redundant or unnecessary information, data aggregation can improve the communication efficiency of a sensor network. A significant risk of data aggregation however is that a node that is captured by an adversary can report arbitrary values as its aggregation result, thereby corrupting not only its own measurements but also that of all the nodes in its entire aggregation sub-tree[6]. As a consequence, an adversary who captures nodes selectively and strategically can corrupt the entire network aggregation process, while incurring minimal cost and effort. Context Summarization (CS) is a method of representing raw context information into summarized information so that it takes relatively less storage space and can successfully answer the queries for complete information with acceptable degree of confidence [7]. Such a compact representation of information reduces required storage space.

It has been identified that police patrols are in need of an integrated version of privacy and context summarization. A typical police patrol system comprises of nodes belonging to police officer, police commissioner and the criminal. The wireless sensors in the network are used to record necessary details. The police commissioner plays a key role in ensuring privacy, performing aggregation and has the access to the database.

## 2. RELATED WORK

A thorough Literature Review of the available papers is done and some of the papers are listed along with the context in which the idea of the paper was studied for the inception of this work.

Computational location privacy algorithms treat location data as geometric information, not as general data [11]. Studies show that people are generally not concerned about location privacy, although they are sensitive to how their location data could be used, and their sensitivity may rise with their awareness of privacy leaks. Offering users to control who can read what tags is difficult due to the high number of items and the lack of user interface [16]. The low resources available on passive RFID tags additionally challenge the use of traditional security protocols. The design of privacy preserving ubiquitous computing context is relative to not only technical implementation but also procedures of services [17].

The proposed way is based that the level of privacy protection ideally should decided by end user of service. Although it has been shown that it presents some flaws and limitations [2], and that finding an optimal k-anonymization is NP-hard, the k-anonymity model is still practically relevant and in recent years a large research effort has been devoted to develop algorithms for k-anonymity.The concept of location k-anonymity for LBS was first introduced in [13] and later extended in [4] to deal with different values of k for different requests. The underlying idea is that a message sent from a user is k-anonymous when it is indistinguishable from the spatial and temporal information of at least k − 1 other messages sent from different users.

The work described in [8] proposes a privacy system that takes into account only the spatial dimension: the area in which location anonymity is evaluated is divided into several regions and position data is delimited by the region. In [3] the concept of mix zones is introduced. A mix zone is an area where the location based service providers can not trace users' movements. When a user enters a mix zone, the service provider does not receive the real identity of the user but a pseudonym that changes whenever the user enters a new mix zone. A similar classification of areas, named sensitivity map is introduced in [14]: it classifies locations as either sensitive or insensitive, and describes three algorithms that hide users' positions in sensitive areas. Moving objects databases (MOD) is another rather young research area that has received a lot of interest in recent years. Several different MOD problems have been tackled, ranging from indexing [15], representing and querying, updating and modelling imprecision and communication costs. Existing work about anonymity of spatio-temporal moving points has been mainly developed in the context of location based services (LBS).

WSNs use multi-hop routing and wireless communication to transfer data and hence are vulnerable to routing attacks. There are a lot of approaches to ease routing security [10]. Some secure AODV algorithms [1] have some effects on defending against external attacks. An on-demand routing protocol for ad hoc to provide resilience to Byzantine failures can be classified into three successive phases: route discovery with fault avoidance, cryptographic primitives and link weight management. An approach to route recovery with one-hop broadcast to bypass compromised nodes in wireless sensor networks was provided by An and Cam[5].

As for multi-path routing, it is more reliable, though it introduces more communication overheads. Multi-path routing, location disguise, and relocation methods can be used to protect base stations [9]. However, in the environment where the network has a large number of compromised nodes, if the compromised node can modify the routing data, system may involve more security issues. PRSA (path redundancy based security algorithm) [21] uses alternative routing paths for each data transmission call to overcome the sensor network attack. To enhance network reliability, PRSA allows sensor node data to be sent on defined routing paths using various transmission modes including round robin, redundant and selective modes.

Fang-Jing Wu and Yu-Chee Tseng [12] defined a data aggregation algorithm that focuses on set of interference neighbors and the transmission directions are toward the sink. But how to schedule multiple tasks at the same time in an efficient way was not discussed. A secure and fault-tolerant data collection scheme [23] with EBS group key management mechanism, termed as CRINet, was proposed, where, the encrypted data reports sent from the source group are relayed to the base station by a 3-way routing approach, for increasing data delivery rate and thus enhancing the fault-tolerant capability. But this increases the communication overhead. Mohanty investigated different possible attacks on the cluster base data gathering protocol and tried to give a symmetric key based security solution [18] to it. This security solution either eliminated or localized the attacks only within a smaller region. Data collection is made secure by using randomized dispersive routes [22], which increases the network overhead.

Bista, Jo, Chang proposed a new and efficient scheme [20] for secure data aggregation where an environment generated sensitive data. The scheme made use of additive property of complex numbers where sensed data were first converted into a complex number form before being transmitted towards the query server [19].

# 3. PPP SYSTEM DESIGN

A system is contrived to achieve Privacy for Police Patrols (PPP). The emplacements of police officers have to be concealed from the criminals. The police commissioner ensures the above and acts as the aggregator to perform context summarization to all the details provided by the police officers.

## 3.1. Flaws in the Existing Methodologies

The existing techniques suffer from several major problems including lack of secrecy and privacy, which makes the network vulnerable to adversaries and attacks. Redundancy exists in data, resulting in data and network overload. Mere XOR operations are performed for data aggregation. Hence, the security factor is not that high.

## 3.2. PPP Proposal

The objective of this work is to develop a system that performs data aggregation in the application of wireless sensor police patrol networks focusing on security and routing to increase the efficiency and reduce the communication overhead. To solve the above described problems, the work centres around a solution that constitutes of location privacy, secure routing, aggregation and pattern identification. This approach integrates the following to ensure privacy, secrecy and less overhead.

### 3.2.1. Location Privacy

Privacy must be insured over the data sent to the nodes in the network. Thus, an efficient algorithm is devised to obfuscate the data to the adversaries and thereby, refraining them from eavesdropping on the content.

### 3.2.2. Secure Routing

It is apparent that, texts can be broken down by the adversaries using cryptanalysis. Hence the process of routing plays an equal role in the process of security. A randomized multi-level routing is formulated to do the necessary.

### 3.2.3. Data Aggregation

The aggregators collect data from a subset of the police officers, aggregate the data using a suitable aggregation function and then transmit the aggregated result to the next module. This is where the redundancy is removed.

### 3.2.4. Pattern Identification

The aggregated data can be summarized by identifying general patterns, which is later used to answer related queries. This adds meaningfulness to the police patrol application.

## 3.3. PPP Architecture

The proposed PPP architecture is illustrated in Figure 1. The RFIDs of the police officers are registered to the Police Commissioner's database in prior. The attacker/criminal attempts to read the position of the nearby police officers. The police commissioner recognizes it as a non-registered ID and replies by obfuscating the positions of the police officers in that zone (Location Privacy module). The police officer sends encrypted data to the police commissioner, which is aggregated to remove redundancy and a pattern, is identified and this knowledge summarized data is stored in the database for further querying.

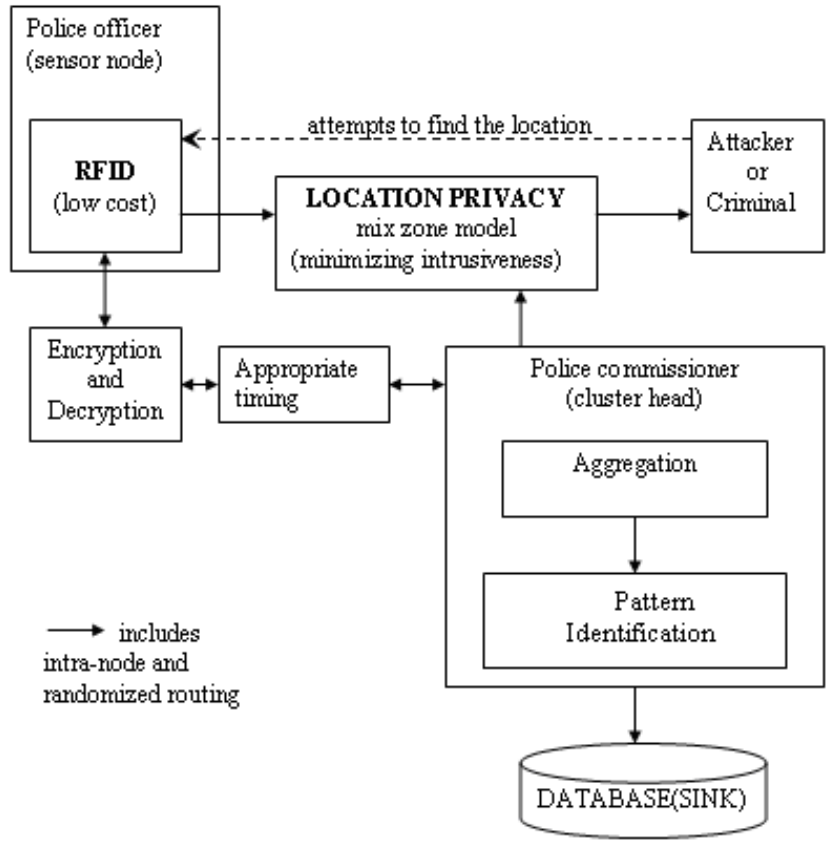

Figure 1. PPP Architecture

### 3.3.1. Location Privacy

Location privacy is needed to ensure that no eavesdropping of critical data takes place. The criteria to achieve privacy are trustworthiness, appropriate timing, perceptibility, unobtrusiveness, minimal intrusiveness, flexibility, meaningfulness and low cost.

Registering each Police Officer with the Police Commissioner in prior ensures the trustworthiness of the system by making it technically reliable and instilling confidence in users. The feedback provided does not distract, annoy or compromise anyone's privacy as it is relevant and selective thereby not overloading the recipient with unwanted information. Identifying the data sent to the Police Commissioner by using the RFID makes the feedback perceptible. Appropriate timing of feedback makes sure that control is provided most likely when needed. The Police Commissioners response immediately follows the reporting of crime by the Police Officer. The system is meaningful with the incorporation of pattern recognition and the cost of design is kept minimal.

Radio Frequency Identification (RFID) Tags represent the most prominent ubiquitous computing technology when it comes to privacy issues. Challenges posed by RFID tags are fourfold. The RFID tags are automated. Data acquisition is made much easier by the use of simple reader gates that can easily scan large numbers of tags. The tag identifies the individual serving its purpose and is well integrated so that no criminal can easily spot it. The last challenge is that posed by authentication as a lot of sensitive information is given.

The four attributes of RFID applications threaten two classes of individual privacy that being data privacy and location privacy. If a tag ID that is associated with a person is spotted at a particular reader location then the location privacy of that person is threatened.

### 3.3.2. Secure Routing

In the PPP system, routing comes into picture while information has to be passed securely between the Police Officer and the Police Commissioner. It is a two way communication process that could occur simultaneously. The routing process is categorized into two levels: intra-node routing and randomized routing.

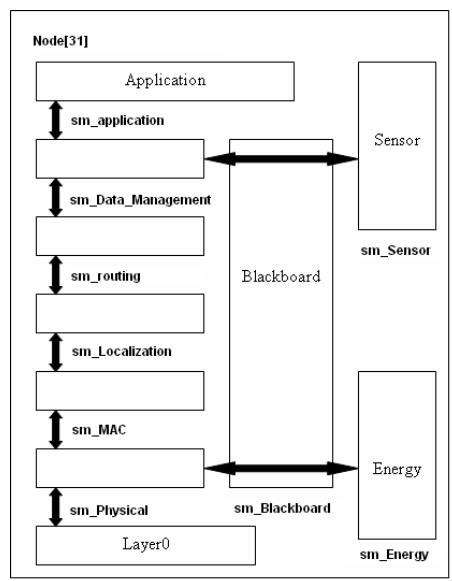

Figure 2. Layered architecture of each node in a PPP system

Each node is fragmented into several layers. The application layer generates the messages and takes care of the debugging. Sensor stores the data in a table (Figure. 2). Data Management does the required aggregation. Localization layer passes the messages to the neighbouring layers. MAC layer reduces the end-to-end packet delay during transmission and also reduces the collisions by putting the data in a queue.

Routing layer implements the randomized routing technique. As for Randomized Routing, the message is split into shares and the message is reconstructed at the aggregator when the minimum needed shares are received.

### 3.3.3. Data Aggregation

Data aggregation is needed in this system so that the database of the Police Commissioner is not overloaded with redundant data of criminals, from the Police Officers. Two or more Police Officers may forward information about the same criminal from the same location of crime to the Police Commissioner. It is enough if only one copy of this information is retained in the database hence saving the storage space and increasing the efficiency of the system.

### 3.3.4. Pattern Identification

A query that needs only limited information may have to process the entire information to find a simple result. This problem is solved by finding meaningful patterns from the information present in the database and storing them in a summarized manner. Hence a query whose answer matches the stored pattern need not perform the time consuming processing.

## 4. IMPLEMENTATION

### 4.1. Location Privacy

The system needs to be protected against adversaries who may attempt to find the location of police officers in the same or nearby zones. The police commissioner takes care of obfuscating the x and y positions of every node inside the zone, thereby ensuring location privacy. This is achieved by a 3 step process given by the algorithm:

*1)* Partitioning
- For all trajectories in D
  - o i <-min T and j<-max T
  - o Insert trajs to D(i,j) s.t i mod   =0 and j mod   =0 and i<=j
  - o D(eq) <- D(i,j) , startT and endT

*2)* Clustering
- Init max_radius
- Repeat
  - o DataSet <-D and Tp<-avg traj
  - o While DataSet not null
    - Tp <- average (max dist (Tp,T))
    - CTp <- Tp and k-1 nearest neighbors
    - If(max dist <max radius)
      - DataSet =DataSet - CTp
      - Clustered = Clustered +CTp
      - Pivot =Pivot + Tp
    - Else Active =Active –Tp

- Follow the same for the left over traj in D else dump it in Trash

*3)* Translation

- Input the anonymity level a
- distortion factor ,df =a/2
- transform every point to a disk of radius, df

In the partitioning step, input dataset D is split into equivalent classes with respect to the timestamp T. The minimum and maximum timestamp of each trajectory is recorded and all those points that lay outside the defined range are discarded. Based on the new starting and ending points, the trajectories are split into equivalent classes D (eq).

As for clustering, the Cluster centers Tp are identified, each one selected farthest from the previous center. From the data set, clusters are formed, each with k trajectories in total, including the center (pivot) and k-1 neighbours. It is also ensured that the cluster radius doesn't exceed the imposed maximum radius. The rest in the trajectory set is discarded.

The ideal aim of the police commissioner is to obfuscate the path traversed by the police officers. Hence, a distortion factor is required to map the original route to a translated route. With an anonymity level a, each point on the clustered trajectory is moved to a uncertainty disk of radius a/2, providing the needed privacy.

## 4.2. Secure Routing

The Routing layer has the required routing information. At layer 0, when a sensor node wants to send a packet to the aggregator, it first breaks the packet into M shares, according to a (T, M)-threshold secret sharing algorithm. Each share is then transmitted to a randomly selected neighbor. That neighbor will continue to relay the share it has received to other randomly selected neighbors, and so on. In each share, there is a TTL field, whose initial value is set by the source node to control the total number of random relays. After each relay, the TTL field is reduced by 1. When the TTL value reaches 0, the last node to receive this share begins to route it toward the aggregator using min-hop routing. Once the aggregator collects at least T shares, it can reconstruct the original packet. No information can be recovered from less than T shares. It then checks for the redundancy of the data before forwarding it to the sink.

## 4.3. Aggregation

The aggregator performs its functionality in the data management layer. Nearly overlapping segments are clustered in each zone into cluster groups, and stored as a single segment per group. The same clustering algorithm as that of the location privacy is implemented here. This method is lossy because it relies on storing summary segments. In the scenario of Police patrols, two or more officers tend to provide the details of the same criminal, belonging to the same zone. Aggregation is performed over the data set and this non redundant data is stored for further queries.

## 4.4. Pattern Identification

Context information can be summarized by identifying general patterns and later answering approximately to the queries using these patterns. From the fed in data, the details of the criminals have been observed. The algorithm followed is discussed in the following steps. The cluster (i.e. zone) is first identified with the minimum and maximum timestamp values and the pivot trajectory. The police officer provides the details of the criminal's location in his input dataset. The close-by segments are clustered using the above devised algorithm. And all the non-registered RFID belonging to the criminal are put up in one file, with the cluster region and zone number. Observing the crime records, crime_number is incremented by 1. The observed pattern is stored in the database for easier and faster query retrieval. Algorithm:

- Identify the zone
- Find the min and max trajs
- Fetch criminal details from the police officers in the corresponding zone
  (File with the format: ID timestamp X Y)
- Cluster the nearly overlapping segments
- From this intermediate file, find
  o criminal ID
  o Crime_number
  o The cluster region
  o zone
- Deposit it in the repository

## 5. RESULT ANALYSIS

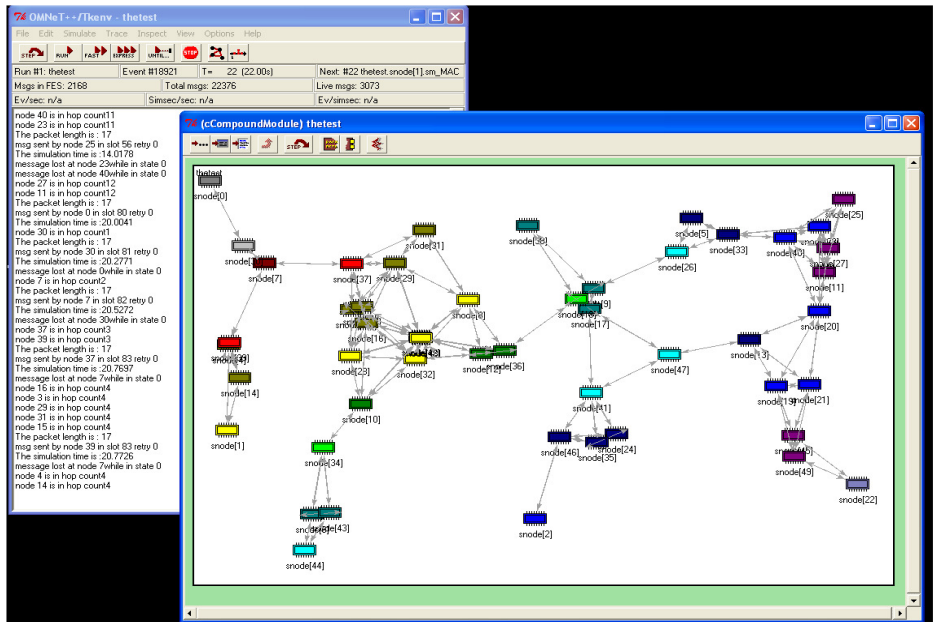

Figure 3. Simulation of PPP Network with 50 nodes.

Figure 3 shows the simulation screenshot of the entire network. The system consists of a network with 50 nodes (police officers and criminals) and aggregator (police commissioner). The colour codes are used to distinguish the zones from each other. The police officers in each zone are clustered and assigned a colour, for visual representation. Aggregated messages are transmitted between the nodes using randomized routing. These unclear messages will avoid the criminal from becoming aware of the police officer's positioning, thereby, achieving location privacy. Every node performs an intra-node routing to ensure proper handling of data.

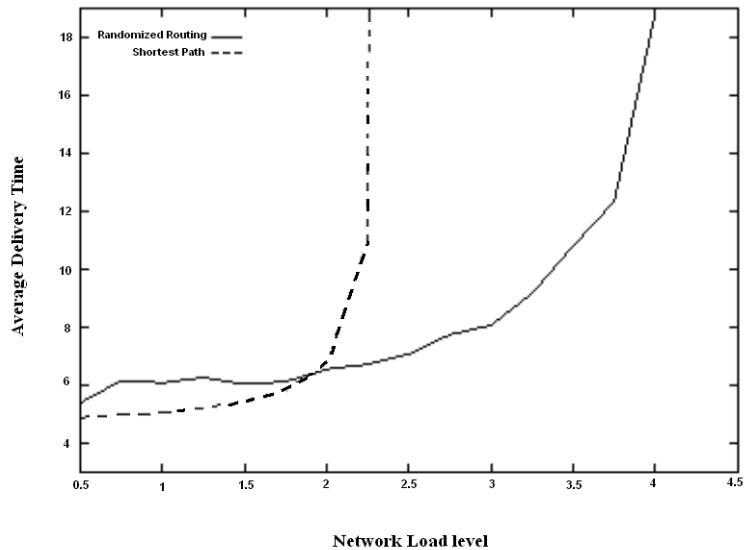

Figure 4. Performance analysis graph to measure the delivery time using this randomized
routing algorithm versus the usual, existing shortest path

The performance analysis graph (Figure 4) shows us that randomized multilevel routing
performs better at higher network load levels when compared to mere shortest path routing. This
proves that when the number of nodes is high; even though the network may seem to be loaded
with obfuscated messages, the messages are delivered efficiently in a faster and a more secure
manner using the randomized routing technique.

```
c:\Users\Uttara\Documents\Visual Studio 2008\Projects\Privacy\Debug\Privacy.exe

Detecting RFID:
Enter the requesting ids:501 502 701 503 702

Police Officers detected:          501        502        503
Criminals detected:        701        702
Choose: 1.Process officers' data         2.obfuscate data to criminals  3.Exit
Choice:1
Enter the zone number:0

Police officers assigned to this zone are: 501  502      503
Enter filename: p1.txt
Enter filename: p2.txt
Enter filename: p3.txt

  Data is aggregated.

Using a context aware pattern, criminals details are recorded.
Press a key.
Choose: 1.Process officers' data         2.obfuscate data to criminals  3.Exit
Choice:2

OBFUSCATING DATA - to achieve LOCATION PRIVACY
@ POLICE COMMISSIONER End:
  Enter the anonymity level and the uncertainity: 57 0.001
Parameters:
  K=57, delta=0.001, pi=5, delta_max=0.010, trash_max=10.0%
Loading data...
  Loading objects... Done.
   -> Trajectories: 1697, Points: 96398, Diameter: 34702.982
   -> Removed Trajectories: 16, Removed Points: 7346
Creating equivalence classes...Done.
Processing equivalence classes: Done!  [ 57 equiv. classes ]

Choose: 1.Process officers' data         2.obfuscate data to criminals  3.Exit
Choice:3
```

Figure 5. Processing results obtained at the police commissioner's end

Figure 5 depicts the processing procedure and results of the PPP system at the Police Commissioner's (PC) site.

The commissioner will initially detect the RFIDs of the incoming requests. Upon identifying the police officers, the police commissioner will process the data fed in. Input files, are collected from the police officers, which has the trajectories of the officer and criminal in the format {RFID, timestamp, x position, y position}. Similar files are collected form the other police officers belonging to the same zone/cluster. Aggregation is performed on the data collected. A required pattern is identified and stored in the database for further query processing.

Any unrecognized RFID is considered to be a criminal. The commissioner will act accordingly and obfuscate the whereabouts of the police officer, using the location privacy technique. The anonymity and the uncertainty levels are set or defined by the police commissioner based on the network load.

## 6. CONCLUSIONS

The overview of the research work carried out in the field of wireless sensor networks is described above and we have listed out the pros and cons of each concept involved.

The problem domain has been analyzed well and a clear outline of the solution has been developed. This work performs the integration of location privacy, routing, a basic aggregation and pattern identification techniques in wireless sensor networks. From the implementation perspective, various privacy algorithms have been studied and implemented; for routing, randomized and multilevel routing are implemented to provide security and efficiency; redundancy has been eliminated by making use of aggregation; and context summarization is performed using pattern identification process for faster query retrieval.

PPP is proved to faster than shortest path routing over a considerably moderate set of nodes. The system can be further extended to work on a larger node set. Future work includes the automation of these integrated mechanisms. Based on the analysis of specific requirements of the police patrol application, we argue that wireless sensor network is very promising in this application and propose an architecture viz. PPP. Also, there are still several research challenges that need to be addressed before the real deployment.